\documentclass[conference]{IEEEtran}
\IEEEoverridecommandlockouts

\usepackage{cite}
\usepackage{hyphenat}
\usepackage{amsmath,amssymb,amsfonts}
\usepackage{subfigure} 
\usepackage{algorithmicx,algorithm}
\usepackage{url}
\usepackage{graphicx}
\usepackage{textcomp}
\usepackage{xcolor}
\usepackage[noend]{algpseudocode}
\usepackage{amsthm}
\usepackage{setspace}
\newtheorem{proposition}{Proposition}

\def\BibTeX{{\rm B\kern-.05em{\sc i\kern-.025em b}\kern-.08em
    T\kern-.1667em\lower.7ex\hbox{E}\kern-.125emX}}

\usepackage{xcolor}
\usepackage{xpatch}
\makeatletter
\def\changeBibColor#1{%
	\in@{#1}{}
	\ifin@\color{blue}\else\normalcolor\fi	
}
\xpatchcmd\@bibitem
{\item}
{\changeBibColor{#1}\item}
{}{\fail}
\xpatchcmd\@lbibitem
{\item}
{\changeBibColor{#2}\item}
{}{\fail}
\makeatother

\begin{document}

\title{Optimizing Fingerprint-Spectrum-Based Synchronization in Integrated Sensing and Communications}

\author{\IEEEauthorblockN{ Xiao-Yang~Wang$^{\dagger \$}$,~Shaoshi~Yang$^\dagger$,~Hou-Yu~Zhai$^\dagger$,~Christos~Masouros$^\$$,~J.~Andrew~Zhang$^\S$}
\IEEEauthorblockA{$^\dagger$School of Information and Communication Engineering, Beijing University of Posts and Telecommnuications, Beijing, China\\
$^\$$Department of Electronic and Electrical Engineering, University College London, London, UK\\
$^\S$Global Big Data Technologies Centre, University of Technology Sydney, Sydney, Australia\\
{E-mails: \{wangxy\_028, shaoshi.yang, 2hy\}@bupt.edu.cn, c.masouros@ucl.ac.uk, andrew.zhang@uts.edu.au}
}

\thanks{
	This work was supported in part by the Beijing Municipal Natural Science Foundation (No. L242013), in part by the National Key R\&D Program of China (No. 2023YFB2904803), in part by the Guangdong Major Project of Basic and Applied Basic Research (No.  2023B0303000001), and in part by BUPT Excellent Ph.D. Students Foundation (No. CX2023238). \textit{Corresponding author: Shaoshi Yang}.
}
}

\maketitle

\begin{abstract}
Asynchronous radio transceivers often lead to significant range and velocity ambiguity, posing challenges for precise positioning and velocity estimation in passive-sensing perceptive mobile networks (PMNs). To address this issue, carrier frequency offset (CFO) and time offset (TO) synchronization algorithms have been studied in the literature. However, their performance can be significantly affected by the specific choice of the utilized window functions. Hence, we set out to find superior window functions capable of improving the performance of CFO and TO estimation algorithms. We first derive a near-optimal window, and the theoretical synchronization mean square error (MSE) when utilizing this window. However, since this window is not practically achievable, we then develop a practical window selection criterion and test a special window generated by the super-resolution algorithm. Numerical simulation has verified our analysis.
\end{abstract}

\begin{IEEEkeywords}
Integrated sensing and communication (ISAC), synchronization, time offset (TO), carrier frequency offset (CFO).
\end{IEEEkeywords}

\section{Introduction}
With the assistance of millimeter wave and terahertz frequencies, high-precision sensing, can be performed in mobile networks \cite{9858656,10226306,9737357,9868166}, resulting in the perceptive mobile network (PMN) concept \cite{zhang2021enabling}. 
PMN relies on a pair of distinct sensing types: passive and active sensing. In passive sensing, the remote radio unit  (RRU) estimates the desired parameters by exploiting the reflected signals of the user equipments (UEs)  or other RRUs. In active sensing, the RRU performs sensing by processing signals transmitted by itself. Nevertheless, high-performance active sensing requires practical full-duplex technology, which is not mature enough to be implemented efficiently at the time of writing \cite{zhang2021enabling}. 
Therefore, passive sensing offers a competitive alternative. 

However, there is a  challenge in achieving high-precision passive sensing. The geographically separated transceivers of passive sensing systems inevitably use different oscillators, which leads to carrier frequency offset (CFO) and time offset (TO) between the transmitters and receivers \cite{10091198}. Furthermore, the CFO and TO will result in nonnegligible velocity and range sensing ambiguity, which severely degrades the sensing performance \cite{ni2021uplink}. 

To mitigate the ambiguity, several kinds of synchronization scheme are proposed \cite{zhang2021enabling}. The first is cross-antenna cross-correlation (CACC)  based algorithms, such as \cite{IndoTrack,ni2021uplink}, while the second is the channel state information (CSI)-ratio based algorithms \cite{FarSense,zeng2020multisense,9904500}. The general framework of these synchronization schemes involves extracting the TO and CFO by processing different signals received from different antennas with the aid of either cross-correlation methods, such as \cite{IndoTrack,ni2021uplink,10616023}, or division operation, such as \cite{FarSense,zeng2020multisense}. However, these research frameworks innately limit applications to the single-antenna and/or non-line-of-sight (NLOS) scenarios \cite{wxy}. 

To broaden the scope of synchronization scenarios, we explore novel insights when designing the cross-multipath cross-correlation (CMCC) synchronization algorithm in \cite{wxy}: 1) the delay-Doppler spectrum of the static signal components, namely the \textit{fingerprint spectrum} \cite{wxy}, is uniquely identified by the  distribution of locations of the static objects reflecting signals; 2) the fingerprint spectrum will cyclically shift, when the CFO and TO drift. As a benefit of these insights, CMCC can estimate high-precision CFO and TO in line-of-sight (LOS)/NLOS and single-target/multi-target scenarios, by locating the peak of the cross-correlation sequence of the fingerprint spectrum at different instants of time. 
Since the fingerprint spectrum corresponds to 
the Fourier transform of the window functions utilized, the window function adopted naturally impacts the estimation performance \cite{wxy}. However, the existing contributions employ standard window functions and there is an open research horizon for re-designing the windows to further enhance the synchronization performance.

In this paper, we conduct theoretical analysis to acquire superior window functions to improve CMCC. 
Firstly, we design an asymptotically optimal window for minimizing the synchronization mean square error (MSE), and derive the corresponding MSE.  
However, since the acquisition of the asymptotically optimal window is practically unachievable, we then design a practical criterion for better window selection.  
Specifically, our key contributions are summarized as follows.
\begin{itemize}
	\item The MSE minimization problem for fingerprint-spectrum-based synchronization is formulated.
	\item[$\bullet$]
	An asymptotically optimal fingerprint spectrum sequence and  its corresponding synchronization MSE are derived.  
\end{itemize}

\begin{itemize}
	\item[$\bullet$]
	Observing the insight: the sharper the main lobe of the Fourier transform of the window peak, the better the synchronization performance, performance of a superior fingerprint spectrum generated by the multiple signal classification (MUSIC) algorithm is tested. 
\end{itemize}

\textit{Notations}:  
${\bf A}^{\textrm{T}}, {\bf A}^{*}, {\bf A}^{\textrm{H}}$ and ${\bf A}^{-1}$ represent transpose, conjugate, conjugate transpose, and inverse of ${\bf A}$, respectively; ${\mathrm{diag}}({\bf a}_1,\cdots,{\bf a}_n)$ is a block{diag}onal matrix whose{diag}onal blocks are $\{{\bf a}_1,\cdots,{\bf a}_n\}$. ${\bf A}[i,:]$ represents the $i$th row of ${\bf A}$; ${\bf a}[i]$ and ${\bf A}[i,j]$ represents the $i$th element of ${\bf a}$ and $(i,j)$th element of ${\bf A}$, respectively; ${\bf I}_N$ and ${\bf 0}_{M\times N}$ are the $N\times N$ identity matrix and $M\times N$ zero matrix, respectively;  $\textrm{Round}(\alpha)$ and $\lfloor\cdot\rfloor$ denote the nearest integer to $\alpha$ and the floor function; Finally, $*$, $\oplus$, and ${\textrm{mod}}(\cdot)$ are the convolution operator, the right cyclic shift operator, and the modulus  operator, respectively.
 
\addtolength{\topmargin}{0.011in}
\section{System Model}

\begin{figure}[tbp]
	\centering
	\includegraphics[width=8.2cm,height=4.1cm]{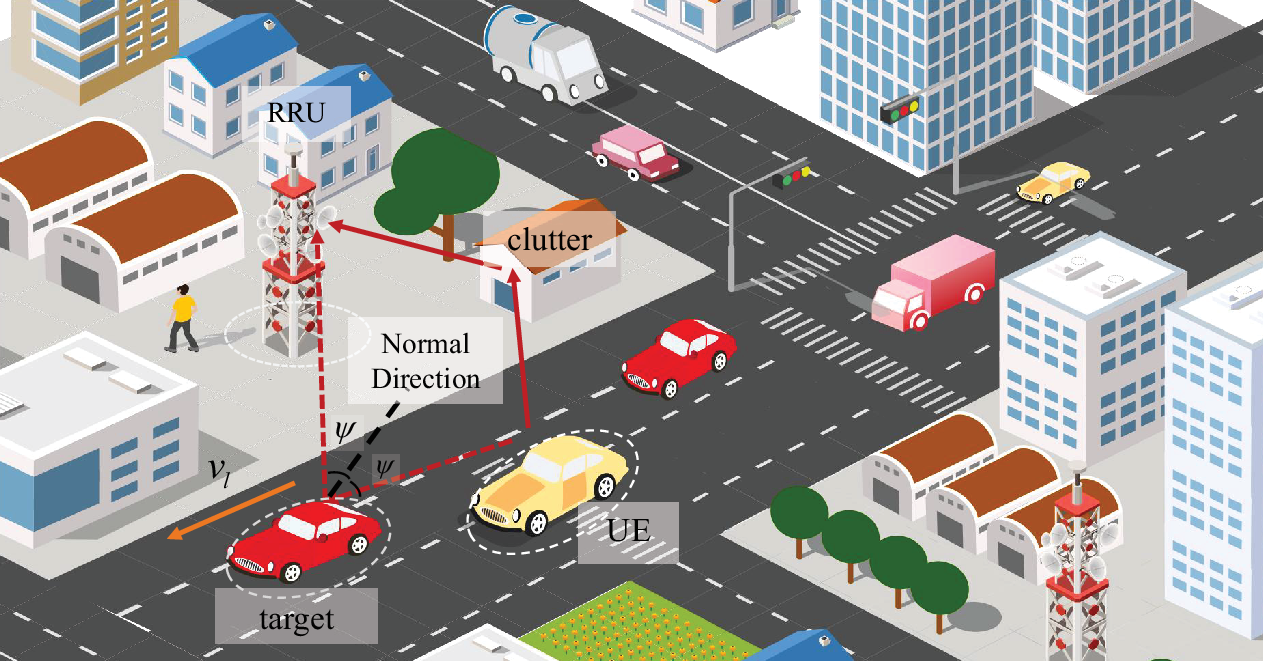}
	\caption{System model.}\label{Sketch}
\end{figure}

The PMN system is depicted in Fig. \ref{Sketch}, where a RRU equipped with an $M_{\textrm{r}}$-element uniform linear array (ULA), receives signals transmitted by the UE. The propagation channel consists of $L$ paths reflected from the objects in the environment. Let us define the propagation delay and the Doppler offset corresponding to the $l$th path as $\tau_{l}$ and $f_{{\textrm{D}},l}$. Then, by estimating $\tau_{l}$ and $f_{{\textrm{D}},l}$, the RRU achieves both velocity measurement and ranging \cite{9921271}. However, the existence of CFO, $\delta^f$, and TO, $\delta^\tau$, between the UE and the RRU, introduce ambiguity in the estimation of $\tau_{l}$ and $f_{{\textrm{D}},l}$ for each path. To mitigate the ambiguity, we assume the whole data payload to be the sensing resource to extract both delay-Doppler and CFO/TO, as described in \cite{zhang2021enabling}.

Specifically, we denote the carrier frequency and the subcarrier spacing in the system by $f_c$ and $\Delta f$. Moreover, let us assume that $N_{\textrm{c}}$ OFDM subcarriers are employed in the system and define $c_n$ as the modulated data symbol of the $n$th subcarrier. 
Additionally, the UE is also equipped with an $M_{\textrm{t}}$-element ULA and the transmit precoder of the UE is denoted as ${\boldsymbol{\omega}}\in\mathbb{C}^{M_{\textrm{t}}\times1}$. Therefore, we can represent the analog signal transmitted by the UE as 
${\bf x}(t)={\boldsymbol{\omega}}e^{j2\pi f_ct}\sum_{n=0}^{N_{\rm c}-1}c_ne^{j2\pi n\Delta ft}$.

Furthermore, the channel impulse response of the $l$th path can be formulated as ${\bf H}_{l}(t)\!=\!a_{l}\delta[(1\!-\!{2v_l}/{c})t\!-\!(\tau_{l}\!+\!\delta^\tau)]{\bf \Omega}_{l}^{\textrm{r}}{\bf \Omega}_{l}^{\textrm{t}}$, 
where $a_{l}$ and $v_l$ represent the channel gain of the $l$th path and the velocity of the reflector projected onto the normal direction in the $l$th path, respectively. Moreover, $\delta(t)$, ${\bf \Omega}_{l}^{\textrm{r}} =[1,e^{j2\pi d\cos(\theta_{l}^{\textrm{r}})/\lambda},\cdots,e^{j2\pi(M_{\textrm{r}}-1) d\cos(\theta_{l}^{\textrm{r}})/\lambda}]^{\textrm{T}}$, and ${\bf \Omega}_{l}^{\textrm{t}} =[1,e^{j2\pi d\cos(\theta_{l}^{\textrm{t}})/\lambda},\cdots,e^{j2\pi(M_{\textrm{t}}-1) d\cos(\theta_{l}^{\textrm{t}})/\lambda}]$ are the impulse function, the received steering vector, and the transmitted steering vector for the $l$th path, respectively. Furthermore, $d$, $\theta_{l}^{\textrm{r}}$, $\theta_{l}^{\textrm{t}}$, and $\lambda$ are defined as the antenna spacing, the angle of arrival (AOA), the angle of departure (AOD) of the $l$th path, and the signal wavelength, respectively. 
Then, the received analog baseband signal of the $g$th OFDM symbol without any cyclic prefix (CP) is formulated as 
\begin{equation}\label{y_g^n(t)}
\begin{aligned}
&{\bf y}_{g}(t) = \ e^{-j2\pi (f_c+\delta^f)t}\ \sum\nolimits_{l=1}^{L}[{\bf H}_{l}(t)\ *\ {\bf x}(t)],
\end{aligned}
\end{equation} where ${\bf w}_{g}(t)$ represents the complex additive white Gaussian noise (AWGN) with zero mean and variance of $\sigma^2$ for any time instant $t$.

To represent the discrete-time signal digitized from ${\bf y}_{g}(t)$, we define the sampling interval, the length of the CP, and the length of an entire OFDM symbol (with CP) as $T_{\textrm{sam}}$, $T_{\textrm{cp}}=N_{\textrm{cp}}T_{\textrm{sam}}$, and $T_{
\textrm{sym}}=N_{\textrm{sym}}T_{\textrm{sam}}$, respectively. 
Then, the $u$th sample of the digitized ${\bf y}_{g}(t)$ including the impact of CP is formulated as 
\begin{equation}\label{Y_g}
\begin{aligned}
{\bf y}_{g,u}\!\!&=\!\!\sum\nolimits_{l=1}^{L}\!\sum\nolimits_{n=0}^{N_{\textrm {c}}-\!1}\!e^{-j2\pi f_c[\frac{2v_l}{c}(uT_{\textrm{sam}}+(g-1)T_{\textrm{sym}}+T_{\textrm{cp}})+(\delta^\tau+\tau_{l})]}\!\\
&\cdot e^{-\!j2\pi n\Delta f\frac{2v_l}{c}(uT_{\textrm{sam}}+(g\!-\!1)T_{\textrm{sym}}+T_{\textrm{cp}})}e^{\!-\!2\pi\delta^f\!(uT_{\textrm{sam}}+(g\!-\!1)T_{\textrm{sym}}+T_{\textrm{cp}})}\\
&\cdot e^{-j2\pi n\Delta f\!(\delta^\tau+\tau_{l})}e^{-j2\pi n\Delta f uT_{\textrm{sam}}}a_{l}c_n{\bf \Omega}_{l}^r{\bf \Omega}_{l}^t{\boldsymbol{\omega}}+{\bf w}_{g,u},
\end{aligned}
\end{equation}
where ${\bf w}_{g,u}$ is defined as the complex AWGN vector with zero mean and covariance matrix $\sigma^2{\bf I}$, respectively. Moreover, noting that $[uT_{\textrm{sam}}\!+\! (g\!-\!1)T_{\textrm{sym}}\!+\!\!T_{\textrm{cp}}]N_{\textrm c}\Delta f\frac{2v_l}{c}$ is always small, one can  approximate  $e^{-j2\pi n\Delta f\frac{2v_l}{c} [uT_{\textrm{sam}}\!+ \!(g-1)T_{\textrm{sym}}+T_{\textrm cp}]}$ by 1 \cite{liu2020super}. Similarly, $e^{-j2\pi f_c[\frac{2v_l}{c} (uT_{\textrm{sam}}+(g-1)T_{\textrm{sym}}+T_{\textrm{cp}})]}$ and $e^{-2\pi\delta^f[uT_{\textrm{sam}} +(g-1)T_{\textrm{sym}}+T_{\textrm{cp}}]}$ can be approximated by $e^{-\!j2\pi f_c[\frac{2v_l}{c}((g-1)T_{\textrm{sym}}+T_{\textrm{cp}})]}$  and $e^{-\!2\pi\delta^f[(g-1)T_{\textrm{sym}}\!+\!T_{\textrm{cp}}]}$, respectively \cite{liu2020super}. 

Consequently, the $g$th OFDM symbol ${\bf Y}_g=[{\bf y}_{g,1},\cdots,{\bf y}_{g,N_{\rm c}}]$ can be  expressed as
\begin{equation}\label{newY_g}
\begin{aligned}
{\bf Y}_g&=\sum\nolimits_{l=1}^{L} e^{-j2\pi f_c\{(\frac{2v_l}{c}+\frac{\delta^f}{f_c})[(g-1)T_{\textrm{sym}}+T_{\textrm{cp}}]\}}e^{-j2\pi f_c(\delta^\tau+\tau_{l})}\\
&\cdot a_{l}{\bf \Omega}_{l}^r{\bf \Omega}_{l}^t{\boldsymbol{\omega}}{\boldsymbol\tau}_{l} {\bf D}{\bf F}\!+\!{\bf W}_g,
\end{aligned}
\end{equation}
where ${\boldsymbol\tau}_{l}=[1,\cdots,e^{-j2\pi (N_{\rm c}-1)\Delta f(\delta^\tau+\tau_{l})}]$, ${\bf D}={\mathrm{diag}}$$([c_0,$ $\cdots,c_{N-1}])$, ${\bf F}$ represents the $N_{\rm c}$-dimensional inverse fast Fourier transform ($N_{\rm c}$-D IFFT) matrix, and ${\bf W}_g=[{\bf w}_{g,1}, \cdots,$ $ {\bf w}_{g,N_{\rm c}}]$, respectively.
To further estimate Doppler and time delay, a common method is to firstly compensate the received signal exploiting the demodulated communication data as $\breve{\bf Y}_g={\bf Y}_g{\bf F}^{\rm H}{\bf D}^{-1}$, which is also exploited in \cite{liu2020super}.

Furthermore, let us denote the $m$th row of $\breve{\bf Y}_g$ as $\breve{\bf y}_{g,m}$. Then, we stack the received signals corresponding $G$ OFDM symbols into $\boldsymbol{\Gamma}_m$ as $[(\breve{\bf y}_{1,m})^{\textrm{T}},\cdots,(\breve{\bf y}_{G,m})^{\textrm {T}}]^{\textrm{T}}$. 
Specifically, there establishes $\boldsymbol{\Gamma}_m=\sum\nolimits_{l=1}^{L}{\boldsymbol\alpha}_{l}[m]\boldsymbol{\theta}_l\boldsymbol{\tau}_l+\breve{\bf W}_m$, 
where ${\boldsymbol\alpha}_{l}=e^{-j2\pi f_c(\delta^\tau+\tau_{l})}a_{l}{\bf \Omega}_{l}^{\textrm{r}}{\bf \Omega}_{l}^{\textrm{t}}{\boldsymbol{\omega}}$, $\boldsymbol{\theta}_l=[e^{-j2\pi f_c(\frac{2v_l}{c}+\frac{\delta^f}{f_c})T_{\textrm{cp}}}, \cdots, e^{-j2\pi f_c(\frac{2v_l}{c}+\frac{\delta^f}{f_c})[(G-1)T_{\textrm{sym}}+T_{\textrm{cp}}]}]^{\textrm{T}}$ and $\breve{\bf W}_m$ is the complex AWGN matrix, respectively.


\section{Windowing for Synchronization and Formulation of Our Optimization Problem}
In this section, we extend the synchronization algorithm in \cite{wxy} by applying window functions for performance improvement and formulate the MSE minimization problem.

As presented in \cite{wxy}, the two-dimensional Fourier transform (2D-FFT) is firstly applied to ${\bar{\boldsymbol{\Gamma}}}_{K,m}$ and $\boldsymbol{\bar\Gamma}_{K,m}$ is a zero-padding-version $\boldsymbol{\Gamma}_m$, which is 
\begin{equation}
\boldsymbol{\bar\Gamma}_{K,m}=\left[
\begin{matrix}
\boldsymbol{\Gamma}_m & {\bf 0}_{G\times (K-1)N_c}\\
{\bf 0}_{(K-1)G\times N_c} & {\bf 0}_{(K-1)G\times (K-1)N_c}
\end{matrix}\right],
\end{equation} where $K$ is the ratio of the number of row/column in $\boldsymbol{\bar\Gamma}_{K,m}$ to that in $\boldsymbol{\Gamma}_m$.
To represent the output of the 2D-FFT, $\boldsymbol{\psi}_{G}\in {\mathbb{C}}^{1\times G}$ is introduced as the $G$-element window function. Then, the output of the 2D-FFT can be expressed as 
\begin{equation}\label{newequ9}
\begin{aligned}
&{\bf F}_{KG}^*{\textrm{diag}}(\boldsymbol{\bar\psi}_{K,G})\boldsymbol{\bar\Gamma}_{K,m}{\textrm{diag}}(\boldsymbol{\bar\psi}_{K,N_c}){\bf F}_{KNc}^*\!=\!\tilde{{\bf W}}_K+\!\sum\nolimits_{l=1}^{L}\!\\
&\boldsymbol{\alpha}_l[m]
[{\bf F}_{KG}^{\textrm{*}}{\textrm{diag}}(\boldsymbol{\bar\psi}_{K,G})\boldsymbol{\bar\theta}_{K,l}][{\bf F}_{KN_c}^{\textrm{H}}{\textrm{diag}}(\boldsymbol{\bar\psi}_{K,N_c}){\boldsymbol {\bar\tau}}^{\textrm{T}}_{K,l}]^{\textrm{T}},
\end{aligned}
\end{equation}
where  $\boldsymbol{\bar\psi}_{K,G}\!=\![\boldsymbol{\psi}_{G},{\bf 0}_{1\times\! (K\!-\!1)G}]$, $\boldsymbol{\bar\psi}_{K,N_c}\!\!=\!\![\boldsymbol{\psi}_{N_c},{\bf 0}_{1\times\! (K\!-\!1)N_c}]$, $\boldsymbol{\bar\theta}_{K,l}\!=\![\boldsymbol{\theta}^{\textrm{T}}_{l},{\bf 0}_{1\times (K\!-\!1)G}]^{\textrm{T}}$, ${\boldsymbol {\bar\tau}}_{K,l}=[{\boldsymbol {\tau}}_{l},{\bf 0}_{1\times (K\!-\!1)N_c}]$, and $\tilde{{\bf W}}_K$ is formulated as
\begin{equation}
\tilde{{\bf W}}_K={\bf F}_{KG}^*\left[
\begin{matrix}
\breve{\bf W}_m & {\bf 0}_{G\times (K-1)N_c}\\
{\bf 0}_{(K-1)G\times N_c} & {\bf 0}_{(K-1)G\times (K-1)N_c}
\end{matrix}\right]{\bf F}_{KN_c}^*.
\end{equation}

Then, we denote ${\bf F}_{KG}^*{\textrm{diag}}(\boldsymbol{\bar\psi}_{K,G})\boldsymbol{\bar\Gamma}_{K,m}{\textrm{diag}}(\boldsymbol{\bar\psi}_{K,N_c}){\bf F}_{KNc}^*$ by $\boldsymbol{\Xi}$, $[{\bf F}_{KG}^{\rm *}{\textrm{diag}}(\boldsymbol{\bar\psi}_{K,G})\boldsymbol{\bar\theta}_{K,l}]$ by the row vector $\boldsymbol{\gamma}_{l,G}$, and $[{\bf F}_{KN_c}^{\textrm{H}}{\textrm{diag}}(\boldsymbol{\bar\psi}_{K,N_c}){\boldsymbol {\bar\tau}}^{\textrm{T}}_{K,l}]^{\textrm{T}}$ by the column vector $\boldsymbol{\gamma}^{\textrm{T}}_{l,N_c}$, respectively.  $\boldsymbol{\gamma}_{l,G}$ is naturally the samples output by the discrete-time Fourier transform (DTFT) of ${\textrm{diag}}(\boldsymbol{\bar\psi}_{K,G})\boldsymbol{\bar\theta}_{K,l}$, which is expressed as
\begin{equation}
\boldsymbol{\bar\gamma}_{l,G}(f)=\sum\nolimits_{g=0}^{KG-1}\boldsymbol{\bar\psi}_{K,G}[g]\boldsymbol{\bar\theta}_{K,l}[g]e^{-j2\pi fg}.
\end{equation}
Similarly, the composition of $\boldsymbol{\gamma}^{\textrm{T}}_{l,N_c}$ is consistent with that of $\boldsymbol{\gamma}_{l,G}$. To associate the continuous DTFT output and the discrete Fourier transform (DFT) output, the sampling interval
has to be determined. According to \cite{wxy,liu2020super}, the sampling interval for $\boldsymbol{\gamma}^{\textrm{T}}_{l,N_c}$ and $\boldsymbol{\gamma}_{l,G}$, namely $T_{\textrm{R}}$ and $F_{\textrm{R}}$, are the the reciprocal of the entire bandwidth, $1/(KN_c\Delta f)$, and the reciprocal of the total time period, $1/(KGT_{\textrm{sym}})$. Thus, we have
$\boldsymbol{\gamma}_{l,G}[i]=\boldsymbol{\bar\gamma}_{l,G}({iF_{\textrm{R}}})=$$\boldsymbol{\bar\gamma}_{l,G}(\frac{i \Delta f}{KG})$ and $
\boldsymbol{\gamma}_{l,N_c}[p]$$=\boldsymbol{\bar\gamma}_{l,N_c}({pT_{\textrm{R}}})$$=\boldsymbol{\bar\gamma}_{l,N_c}(\frac{pT_{\textrm{sam}}}{K}).
$
As a result, $\boldsymbol{\Xi}[i,p] $ can be expressed as
\begin{equation}\label{xi}
\begin{aligned}
\boldsymbol{\Xi}[i,p]\!\!=\!\! \sum_{l=1}^{L}2\pi^2\boldsymbol{\alpha}_l[m]\boldsymbol{\bar\gamma}_{l,N_c}\!\left(\!\frac{pT_{\textrm{sam}}}{K}\right)\!\boldsymbol{\bar\gamma}_{l,G}\!\left(\!\frac{i \Delta f}{KG}\right)\!\!+\!\!{\tilde{\bf W}}_K[i,p].
\end{aligned}
\end{equation} 

Therefore, according to definition of $\boldsymbol{\gamma}_{l,G}$ and $\boldsymbol{\gamma}_{l,N_c}$, it is intuitive that $\boldsymbol{\Xi}$ is the sum of the Gaussian noise and the linear summation of $L$ discrete 2-D sample matrices, namely $\boldsymbol{\gamma}_{l,G}\boldsymbol{\gamma}^{\textrm{T}}_{l,N_c}$, whose centers are at $[(\delta^f+f_{{\textrm{D}},l})/F_{\textrm{R}} ,(\delta^\tau+\tau_l)/{T_{\textrm{R}}}]$ for $l=1,\cdots,L$. Moreover, each 2D sample matrix, $\boldsymbol{\gamma}_{l,G}\boldsymbol{\gamma}^{\textrm{T}}_{l,N_c}$, corresponds to signals  reflected by a object. Among the $L$ objects, we assume that there exist $L_{\textrm{s}}$ static objects. 
It is natural that the symmetry points of these $L_{\textrm{s}}$ matrices are $(\delta^f/F_{\textrm{R}},(\delta^\tau+\tau_l)/{T_{\textrm{R}}})$ because $f_{{\textrm{D}},l}=0$ for these static objects, which means that symmetry points of these 2D matrices are located at the same row index. 

By locating the row index of the $L_{\textrm{s}}$ symmetry points, which is assumed as $K_0$, we find the \textit{fingerprint spectrum} sequence ${\boldsymbol{\beta}}_{\textrm{n}}\!\!\! =\!\!\boldsymbol{\Xi}[K_0,:]$. 
Essentially, the sequence may be deemed to be an identity code of the specific static environment. 
However, ${\boldsymbol{\beta}}_{\textrm{n}}$ is the fingerprint spectrum polluted by noise. For simplicity, we here define ${\boldsymbol\beta}\in\mathbb{C}^{1\times KN_c}$ as the pure fingerprint spectrum vector without noise and ${\boldsymbol{\beta}}$ can be formulated as ${\boldsymbol{\beta}}={\boldsymbol{\beta}}_{\textrm{n}}-\tilde{\bf W}_K[K_0,:]$.

Then, if the CFO increases by $\Delta\delta^f$ and the TO increases by $\Delta\delta^\tau$, the offset fingerprint spectrum contaminated by noise can be defined as ${\boldsymbol{\beta}}_{\textrm{n, c}} \!\!=\!\boldsymbol{\Xi}[K_0+{\textrm{Round}}(\Delta\delta^f/F_{\textrm{R}}),:]$ according to (\ref{xi}). Moreover, we also define ${\boldsymbol{\beta}}_{\rm c}$ as the offset fingerprint spectrum without noise. Therefore, since ${\textrm{Round}}(\Delta\delta^f/F_{\textrm{R}})\approx\Delta\delta^f/F_{\textrm{R}}$ due to large $K$ and $N_c$, the relationship between ${\boldsymbol{\beta}}$ and ${\boldsymbol{\beta}}_{\rm c}$ can be simply represented as ${\boldsymbol{\beta}}_{\rm c}[q]\approx{\boldsymbol{\beta}}[q+\frac{\Delta\delta^\tau}{T_{\textrm{R}}}]$ \cite{wxy}.
As a result, in order to find the number of cyclic shifts and consequently estimate the CFO and TO, the sliding-window cross-correlation analysis between ${\boldsymbol{\beta}}_{\textrm{n}}$ and $\boldsymbol{\Xi}$ can be implemented as \cite{wxy}
\begin{equation}\label{unsimplified}
\begin{aligned}
\{\!&{\Delta\hat\delta^f}\!/{F_{\textrm{R}}},{\Delta\hat\delta^\tau}\!/{T_{\textrm{R}}}\!\}\!=\!\mathop{\textrm{max}}\limits_{i,q}\!\big|\!\sum\nolimits_{p=1}^{{KN_c}}\!{{\boldsymbol{\Xi}}[i,\bar q]{\boldsymbol{\beta}_{\textrm{n}}}^*[p]}/{|{\boldsymbol{\beta}_{\textrm{n}}}^*|^2}\big|,\\
\end{aligned}
\end{equation}
where $\bar q=[(q+p)\ {\textrm{mod}}\ {KN_c}]$. 

For simplicity, let us define  $\sum_{p=1}^{{KN_c}}{{\boldsymbol{\Xi}}[i,\bar q]{\boldsymbol{\beta}_{\textrm{n}}}^*[p]}/{|{\boldsymbol{\beta}_{\textrm{n}}}^*|^2}$ as ${\bf A}[i,q]$, and by substituting (\ref{xi}) and ${\boldsymbol{\beta}_{\textrm{n}}}$ into (\ref{unsimplified}), ${\bf A}[i,q]$ can be further represented as
\begin{small}
	\begin{equation}\label{correlation}
	\begin{aligned}
	&{\bf A}[i,q]\!\!=\! 4\pi^4\!\frac{{\boldsymbol\gamma}_{1,G}\!\left[i\right]\!{\boldsymbol\gamma}_{1,G}\!\left[\! K_0\!\right]}{{|{\boldsymbol{\beta}_{\textrm{n}}}^*|^2}}\!\sum_{l^{'}=1}^{L}\!\sum_{l=1}^{L}\boldsymbol{\alpha}_l[m]\boldsymbol{\alpha}_{l^{'}}[m]{\boldsymbol\rho}_{l,l^{'}}[q]\!+\!{\bf \bar W},
	\end{aligned}
	\end{equation}\end{small}where ${\bf \bar W}$ is the complex AWGN matrix with zero mean and variance $\bar\sigma^2$. Furthermore, ${\boldsymbol\rho}_{l,l^{'}}\in \mathbb{C}^{1\times KN_c}$ is formulated as
\begin{small}\begin{equation}\label{rho}
	{\boldsymbol\rho}_{l,l^{'}}[q]\!=\!\sum\nolimits_{p=1}^{{KN_c}}\!\!{{\boldsymbol{\bar\gamma}}_{l,N_c}\!\left[[(p+q)\ {\textrm{mod}}\ KN_c]T_{\textrm{R}}\right]\!{\boldsymbol{\bar\gamma}}_{l^{'}\!,N_c}\left[pT_{\textrm{R}}\right]}.
	\end{equation}\end{small}

From (\ref{correlation}) and (\ref{rho}), we find that ${\bf A}[i, q]$ is strongly coupled with ${\boldsymbol\gamma}^{\textrm{T}}_{l,N_c}$ for $l=1,\cdots,L$, while ${\boldsymbol\gamma}_{l,G}$ only affects the signal-to-noise ratio (SNR) of ${\bf A}[i, q]$. 
Thus, the impact of the row index, $k$, on the estimation performance is equivalent to that of the SNR. As a result, we here set $k$ as a constant $K_1={\textrm{Round}}(\Delta\delta^f/F_{\textrm{R}})$, which makes ${\boldsymbol{\gamma}}_G\left[ K_1\right]{\boldsymbol{\gamma}}_G\left[K_0\right]$ as large as possible for simplicity. 

As a result, the MSE of the delay estimate in (\ref{unsimplified}) is a function of $\boldsymbol{\psi}_{N_c}$
\begin{equation}
f_{\textrm{MSE}}(\boldsymbol{\psi}_{N_c}) = \sum\nolimits_{q=1}^{KN_c}{\rm P}_{\{q,KN_c\}}(q-{\Delta\delta^\tau}/{T_{\textrm{R}}})^2,
\end{equation}   
where ${\rm P}_{\{q,KN_c\}}$ is the probability that the $q$th element of the set $\{{\bf A}[K_1,q]| q=1,\cdots,K{N_c}\}$ is the maximum. According to the definition, ${\rm P}_{\{q,KN_c\}}$ can be computed as the cumulative distribution function (CDF) of a $KN_c$-dimensional Gaussian random vector, yielding
\begin{equation}\label{newequ15}
\begin{small}
	\begin{aligned}
	&{\rm P}_{\{q,KN_c\}} = \int_{\mathbb{V}}p({\bf s},\tilde\sigma^2){\mathrm d}\ {\bf b}\\
	&=\int_{-\infty}^{\infty} \underbrace{\int_{-\infty}^{b_q}\cdots\int_{-\infty}^{b_q}}_{KN_c-1} \frac{e^{-{ ({\bf b}-{\bf s})^{\textrm{T}}({\bf b}-{\bf s})}/{2\tilde\sigma^2}}}{(2\pi)^{\frac{d}{2}}\tilde\sigma^{d}}{\mathrm d}\ b_1\cdots {\mathrm{d}}\ b_{KN_c}{\mathrm{d}}\ b_q,
	\end{aligned}
	\end{small}
\end{equation}where $p({\bf s},\tilde\sigma^2)$ is the probability of the multi-dimensional Gaussian function with mean ${\bf s}$ and covariance matrix $\tilde\sigma^2{\bf I}_{KN_c}$. Furthermore,  ${\bf s}$ is the expectation of the $K_1$th row of ${\bf A}$. Note that A is the linear weighted sum of the circular convolution of ${\boldsymbol{\gamma}}_{N_c}$ according to (\ref{correlation}) and (\ref{rho}), while ${\boldsymbol{\gamma}}_{N_c}$ is the Fourier transform of the selected window function. As a result, the window function $\boldsymbol{\psi}_{N_c}$ has a strong influence on the synchronization MSE. 
Moreover, ${\bf b}=(b_1,\cdots, b_{KN_c})$ in (\ref{newequ15}) belongs to the $KN_c$-element vector space $\mathbb V$, which satisfies $b_q>b_i, {\rm for}\ i\in\{1,\cdots,KN_c\}\ {\rm and}\ i\neq q$.



In the following, we intend to optimize the window function to minimize $f_{\textrm{MSE}}(\boldsymbol{\psi}_{N_c})$:
\begin{equation}
\label{P}
\begin{aligned} 
\mathop{\rm min}\limits_{\boldsymbol{\psi}_{N_c}}\ & f_{\textrm{MSE}}(\boldsymbol{\psi}_{N_c}), \\
{\rm s.t.}\quad 
&\boldsymbol{\psi}_{N_c}\in\mathbb{C}^{1\times KN_c}.
\end{aligned}
\end{equation}

\section{Asymptotically Optimized Ideal Window Function}
We intend to find the solutions of (\ref{P}) by the following logic.
Firstly, in Proposition 1, we derive an asymptotically optimal solution for the following problem
\begin{equation}
\begin{small}
	\label{P1}
	\begin{aligned} 
	\mathop{\textrm{min}}\limits_{{{g }(\boldsymbol{\psi})}}\ & f_{\textrm{MSE}}({{g }(\boldsymbol{\psi})}), \\
	{\textrm{s.t.}}\quad 
	&{{g }(\boldsymbol{\psi})}\in\mathbb{C}^{1\times KN_c}, {\boldsymbol{\psi}\in\mathbb{C}^{1\times KN_c},}
	\end{aligned}
	\end{small}
\end{equation}where ${g }(\boldsymbol{\psi})$ is defined as the function with respect to the window function $\boldsymbol{\psi}$, namely ${\bf s}$. Moreover, Proposition 1 also derives the MSE corresponding to the asymptotic optimal solution.
	Then, based on the relation between ${{g }(\boldsymbol{\psi})}$ and $\boldsymbol{\psi}$, Proposition 2 
formulates the cross-correlation of the Fourier transform of window function corresponding to specific fingerprint spectrum. Finally, based on Proposition 2, Proposition 3 specifies the approximately optimal  $\boldsymbol{\psi}_{N_c}$ corresponding to any ${\bf s}$. 

Concretely, Proposition 1 is described as:

\begin{proposition}
	If $\frac{\Delta\delta^\tau}{T_{\textrm{R}}}$ is an integer, $f_{\textrm{\rm{MSE}}}(\boldsymbol{\psi}_{N_c})$ is asymptotically  minimized when the SNR tends to infinity and ${\bf s}$ satisfies
	\begin{equation}\label{ideal}
	\begin{cases}
	{\bf s}[q]=1, q={\Delta\delta^\tau}/{T_{\textrm{R}}},\\
	{\bf s}[q]=0, q\neq{\Delta\delta^\tau}/{T_{\textrm{R}}},
	\end{cases}
	\end{equation} 
	and the corresponding MSE is  $f_{\textrm{\rm{MSE}}}(\boldsymbol{\psi}_{N_c})=\sum_{\substack{q=1, q\neq {\Delta\delta^\tau}\!/{T_{\textrm{R}}}}}^{KN_c} $ $[ q_{\textrm{\rm{r}}}\!-\! \frac{\Delta\delta^\tau}{T_{\textrm{R}}}]^2$$\int_{-\infty}^{\infty} \!\big[f_{\mathcal N}(b_q)Q(\frac{b_q-1}{\bar\sigma^2})Q^{(KN_c-2)} ({b_q}/{\bar\sigma^2})\big]{\mathrm d} b_q$, where $Q(\cdot)$ is the cumulative distribution function of the standard Gaussian random variable.
	\label{pro1}
\end{proposition}

\textit{Proof}: Due to the page limit, please refer to \cite{10640151}.

However, unlike the hypothesis of Proposition \ref{pro1}, $\frac{\Delta\delta^\tau}{T_{\textrm{R}}}$ can hardly be an integer in actual PMN systems. According to the criterion described in proof of Proposition 1, to make  $f_{\textrm{MSE}}(\boldsymbol{\psi}_{N_c})$ as small as possible, one should assign as much possibility as possible to the index closest to $\frac{\Delta\delta^\tau}{T_{\textrm{R}}}$ and as low possibility as possible to indices far from $\frac{\Delta\delta^\tau}{T_{\textrm{R}}}$ as the maximum value. Under this case, the probability corresponding to ${\textrm{Round}}(\frac{\Delta\delta^\tau}{T_{\textrm{R}}})$ should be assigned as large weight as possible and there holds
\begin{equation}\label{s_q}
\begin{cases}
{\bf s}_{\textrm{ap}}[q]=1, q={\textrm{Round}}(\frac{\Delta\delta^\tau}{T_{\textrm{R}}}),\\
{\bf s}_{\textrm {ap}}[q]=0, q\neq{\textrm{Round}}(\frac{\Delta\delta^\tau}{T_{\textrm{R}}}),
\end{cases}
\end{equation}
Similarly to the case where $\frac{\Delta\delta^\tau}{T_{\textrm{R}}}$ is an integer, ${\bf s}_{\textrm{ap}}$ in (\ref{s_q}) will be asymptotically optimal, when $1/\tilde{\sigma}^2$ becomes large.
Specifically, based on (\ref{newequ15}), under this case we further derive ${\rm P}_{\{q,KN_c\}}$ by dividing $q=1,\cdots,KN_c$ into two categories, namely $q={\textrm{Round}}(\frac{\Delta\delta^\tau}{T_{\textrm{R}}})$ and $q\neq{\textrm{Round}}(\frac{\Delta\delta^\tau}{T_{\textrm{R}}})$
\begin{equation}\label{equ28}
\begin{cases}
\int_{-\infty}^{\infty}f_{\mathcal N}({b_q}/{\bar\sigma^2})Q^{(KN_c-1)}({b_q}/{\bar\sigma^2}){\mathrm d}b_q,\ q={\textrm{Round}}(\frac{\Delta\delta^\tau}{T_{\textrm{R}}}),\\
\int_{-\infty}^{\infty}\!\!f_{\mathcal N}(b_q)Q(\frac{b_q-1}{\bar\sigma^2})Q^{(KN_c\!-\!2)}({b_q}/{\bar\sigma^2}){\mathrm{d}} b_q,q\!\neq\!{\textrm{Round}}(\frac{\Delta\delta^\tau}{T_{\textrm{R}}}).
\end{cases}
\end{equation}
Furthermore, according to (\ref{equ28}), $f_{\textrm{MSE}}$ for non-integer $\frac{\Delta\delta^\tau}{T_{\textrm{R}}}$ can be simply expressed.

Although Proposition 1 describes the near-optimal ${\bf s}$ and the corresponding $f_{\textrm{MSE}}(\boldsymbol{\psi}_{N_c})$, we have not acquired the expression of ${\boldsymbol\rho}_{l,l^{'}}$ corresponding to the near-optimal ${\bf s}_{\rm ap}$. Essentially, ${\bf s}_{\rm ap}$ is constituted by the circular auto-correlation sequence of the discrete Fourier transform of the selected window functions, namely ${\boldsymbol\rho}_{l,l^{'}}$ for $l, l^{'}=1,\cdots\!,L$. Moreover, 
for any $l$ and $l^{'}$, ${\boldsymbol\rho}_{l,l^{'}}$ is a circular-shifted version of ${\boldsymbol\rho}_{l,l}$ and for any $l$ we have: ${\boldsymbol\rho}_{1,1}={\boldsymbol\rho}_{l,l}$ according to (\ref{rho}). Thus, we further derive the exact ${\boldsymbol\rho}_{1,1}$ in Proposition \ref{pro2}, which outlines the closed-form expression of the desired ${\boldsymbol\rho}_{1,1}$ as

\begin{proposition}\label{pro2}
$\boldsymbol{\rho}_{1,1}$ corresponding to any achievable ${\bf s}$ is $\boldsymbol{\rho}_{1,1}={{\bf s}\breve{\bf J}_{\textrm{\rm{Inv}}}}$, where $\breve{\bf J}_{\textrm{\rm{Inv}}}={\bf F}{\rm{diag}}( {\breve{\boldsymbol{\phi}}}_{\rm F}^{\textrm{\rm{Inv}}}){\bf F}^{\rm H}$ and ${\breve{\boldsymbol{\phi}}}_{\rm F}^{\textrm{\rm{Inv}}}=\tilde{\boldsymbol \phi}_m{\bf J}{\bf F}$. 
\end{proposition}

\textit{Proof}: Due to the page limit, please refer to \cite{10640151}.

Specifically, $\tilde{\boldsymbol \phi}_m$ is then acquired by rearranging $[\boldsymbol{\alpha}_1[m]\boldsymbol{\alpha}_{1}[m]$$,\cdots,\boldsymbol{\alpha}_l[m]\boldsymbol{\alpha}_{l^{'}}[m],\cdots,\!\boldsymbol{\alpha}_L[m]\boldsymbol{\alpha}_{L}[m]]$ in ascending order of ${\Delta_{l,l^{'}}}=(\tau_l-\tau_{l^{'}}) \ \textrm{mod} \ KN_c$. ${\bf J}$ is the circulant matrix
\begin{equation}\label{J}
	{\bf J}=
	\begin{pmatrix}
	&{\bf 0}_{KN_c-1} &{\bf I}_{KN_c-1}\\
	&1 &{\bf 0}_{KN_c-1}
	\end{pmatrix}.
\end{equation}

According to this proposition, the value of the global optimum ${\boldsymbol \psi_{N_c}}$ will be strongly related to complex gain and time delay of every path, which is hard to acquire before successful sensing. Thus, even though there exists a globally optimal $\boldsymbol{\psi}_{N_c}$, we cannot practically find it since it is strongly time-delay-and-complex-gain-dependent, before we obtain the time delay and complex gain of all paths.

\section{Practical Window Selection Criterion}

Even though the asymptotically optimal window cannot be practically deployed, a large number of feasible discrete windows belonging to $\mathbb{C}^{1\times KN_c}$ can be selected. However, there are no criteria clarifying how to choose good windows from the set of the massive variety of feasible windows $\boldsymbol{\psi}_{N_c}$. 

To establish a criterion, we revisit Proposition \ref{pro1}. According to this proposition, the cross-correlation sequence of the fingerprint spectrum follows a complex Gaussian distribution with a mean of $\sum_{p=1}^{KN_c}{\boldsymbol{\beta}}[p]{\boldsymbol{\beta}}_{\textrm{c}}[p+q]$ for $q=1,\cdots\!,KN_c$, and a variance of $\bar\sigma^2$. Additionally, when $\tilde\sigma^2$ is known, $\bar\sigma^2$ is determined by $|{\boldsymbol{\beta}_{\textrm{n}}}|^2$.

\begin{figure}[tbp]
	\centering
	\includegraphics[width=6.6cm,height=3.5cm]{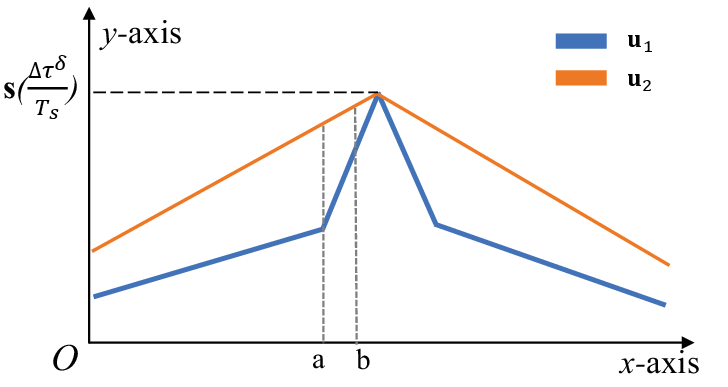}
	\caption{${\bf u}_1$ and ${\bf u}_2$ are two implementations of ${\bf s}$. ${\bf u}_1$ has a sharper mainlobe than ${\bf u}_2$.}
	\label{Pro1}
\end{figure}

In practical  high-SNR scenarios, the noise power $\bar\sigma^2$ is significantly smaller than the difference between two consecutive elements on the main lobe of the mean sequence, ${|{\boldsymbol{\beta}_{\textrm{n}}}|^2}{\bf s}$, such as ${|{\boldsymbol{\beta}_{\textrm{n}}}|^2}[{\bf s}({\textrm{Round}}(\frac{\Delta\delta^\tau}{T_{\textrm{R}}}))-{\bf s}({\textrm{Round}}(\frac{\Delta\delta^\tau}{T_{\textrm{R}}}-1))]$.
In such cases, if the main lobe becomes narrower, although $|{\boldsymbol{\beta}_{\textrm{n}}}|^2$ may decrease, the change in $\bar\sigma^2$ is relatively smaller than the change in the difference between two adjacent elements. As a result, as shown in Fig. \ref{Pro1}, the probability that ${\bf u}_2(a)$ is higher than ${\bf u}_2(b)$ will be lower than the probability that ${\bf u}_1(a)$ is higher than ${\bf u}_1(b)$. Consequently, the maximum will be located closer to the peak point of ${\bf s}$, when ${\bf s}$ has narrower lobe. 

As (\ref{correlation}) shows, ${\bf s}$ is the linear weighted sum of the cyclic shifted $\boldsymbol{\rho}_{1,1}$.
Moreover, as (\ref{correlation}) suggests, $\boldsymbol{\rho}_{1,1}$ can be represented as the convolution of the Fourier transforms of the windows utilized and the reversed sequences of the Fourier transforms, which is equivalent to the autocorrelation of the Fourier transform of the  window used. In accordance with the definition of the autocorrelation, it can be viewed as the sliding multiply and sum operation for a certain sequence. Then, it is intuitive that the mainlobe sharpness of the cross-correlation, namely $\boldsymbol{\rho}_{1,1}$, is related to the mainlobe sharpness of the Fourier transform of the window utilized, given the usually high attenuation of the sidelobes. As a result, the sharper the mainlobe of the Fourier transform of the utilized window peak is, the better the estimation performance.

\textit{Based on the above insight, we aim to find windows whose Fourier transform has a sharp mainlobe to advance the estimation performance.} Obviously, there are no other traditional windows whose Fourier transform has a narrower mainlobe width than that of the rectangular window. Therefore, the rectangular window has the best performance among the traditional windows. Moreover, to further verify the finding, we utilize the super-resolution estimation algorithm (SREA) to generate special windows with extremely narrow mainlobe and test its performance. Generally, we process ${\boldsymbol{\Gamma}_m}$ by replacing the traditional 2D-FFT in (\ref{newequ9}) by the MUSIC algorithm. The outputs of the MUSIC algorithm are  bell-shaped curves. If the MUSIC algorithm is regarded as a spectral analysis tool, like DFT, then the outputs of the MUSIC can be accordingly viewed as spectral components filtered by the ``bell-shaped window", assuming the constant shape of the window. Thus, the ``bell-shaped window" is treated as a special window generated by the MUSIC algorithm. Given the generally sharp peak of the bell-shaped curve \cite{6942180}, the corresponding performance will be improved.

\section{Numerical Simulations}

\begin{figure}[tbp]
	\centering
	\subfigure[]{
		\vspace{-1.5cm}
		\includegraphics[width=0.48\linewidth]{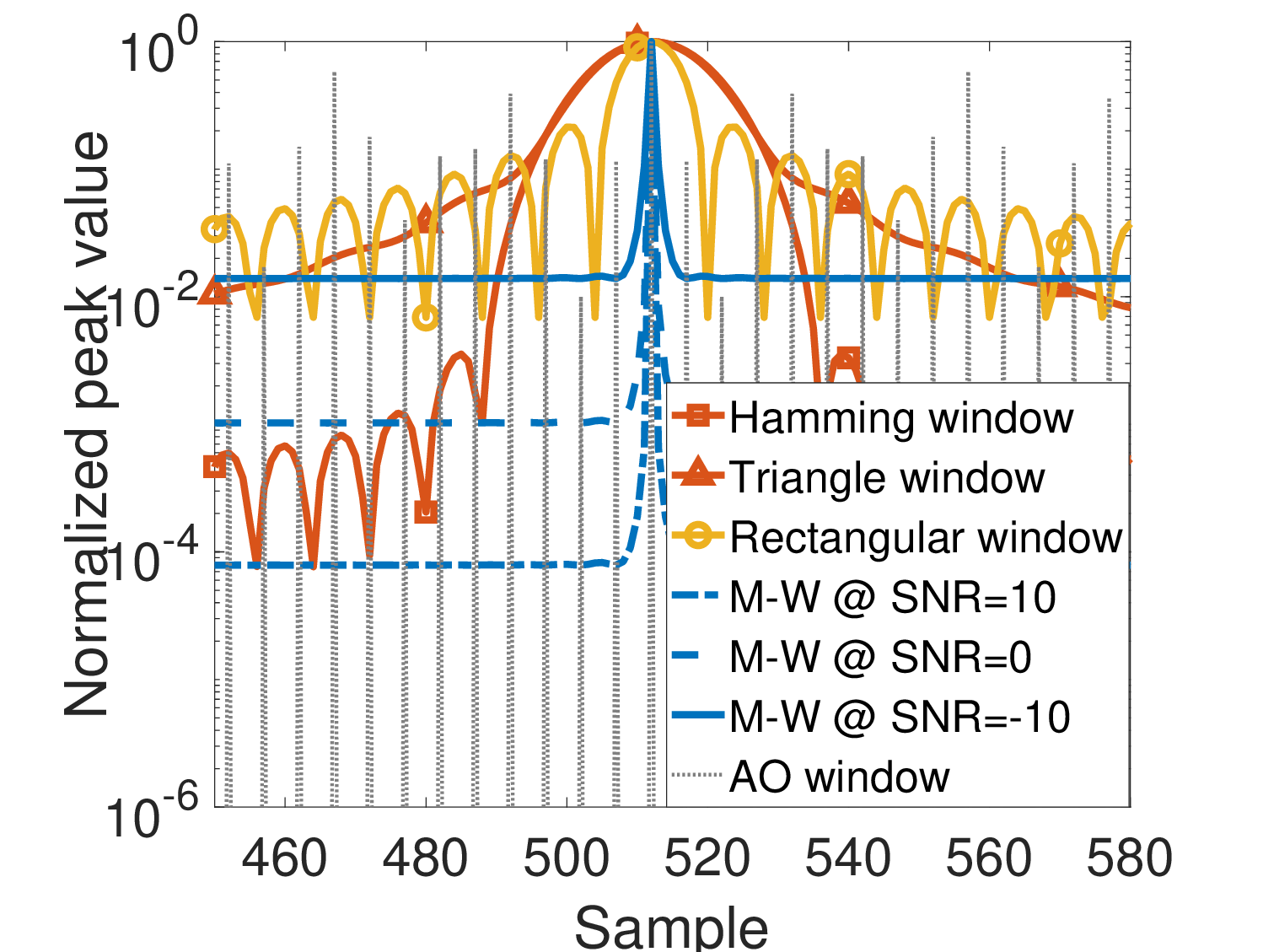}}\subfigure[]{
		\includegraphics[width=0.48\linewidth]{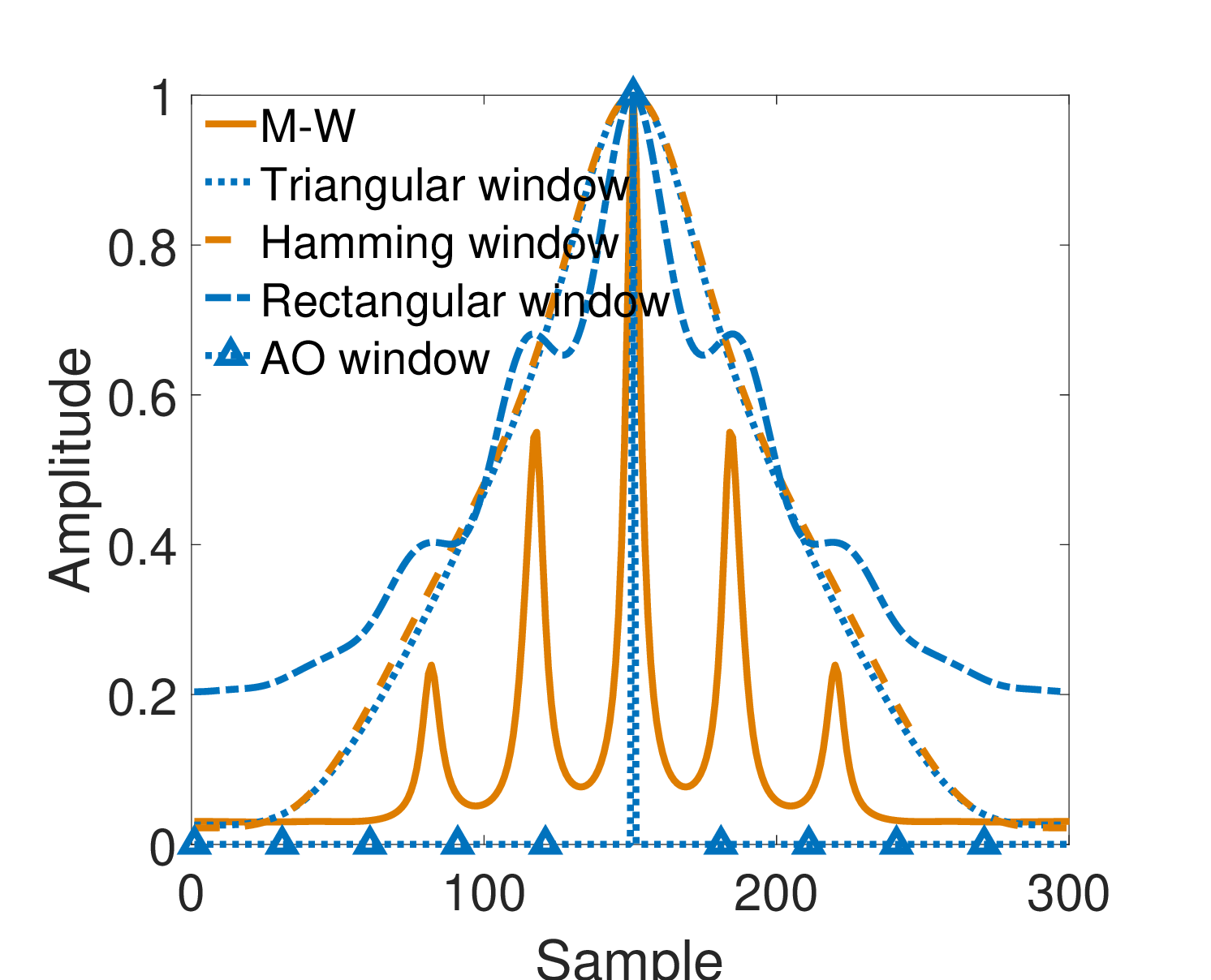}}
	\caption{(a): Auto-correlation of the MUSIC window, Fourier transforms of different traditional windows, and the asymptotically optimal window. (b): The auto-correlation sequence of fingerprint spectrum corresponding to different windows. ``M-W" and ``AO window" represent correlation of the bell-shaped window and our proposed asymptotically optimal window, respectively. 
	}\label{Correlation_fingerprint}
\end{figure}

In this section we perform numerical simulations for verifying our theoretical analysis. In the simulations, we assume that the number of RRU and UE antennas are $M_{\rm r}=64$ and $M_{\textrm{t}}=2$, respectively. The carrier frequency $f_c$ is assumed to be $28$ GHz and the subcarrier spacing is set to $100$ KHz \cite{ni2021uplink}. Moreover, we assume that the number of subcarriers is $128$ and the length of CP is $16$.

In subfigure (a) of Fig. \ref{Correlation_fingerprint} we draw the auto-correlation sequence of the Fourier transforms of different windows, including the traditional windows, the bell-shaped window and the near-optimal window. Subfigure (b) draws the cross-correlation sequence of the fingerprint spectrum generated by the traditional windows, bell-shaped windows and the asymptotically optimal window.  
According to the subfigure (a), although the noise floor of the bell-shaped window varies when SNR and the number of snapshots increase, the width of the mainlobe of the window is invariably lower than that of all the traditional windows. Moreover, since the asymptotically optimal window shape is dependent on a specific scenario, we draw its correlation in the same scenarios as those in subfigure (b) for intuitive comparison. 
As the subfigure (b) shows, the fingerprint spectrum generated by the bell-shaped window exhibits sharper peaks than the mainlobe of the traditional windows,  while the peak of the asymptotically optimal window is the sharpest. Actually, the shape of ``The AO window" curve is formulated as (\ref{s_q}). As the criterion described in Section \uppercase\expandafter{\romannumeral5} shows, the synchronization performance for the AO window will outperform the other windows, while the performance for the MUSIC window will outperform the traditional windows. This conclusion only holds if the shape of the bell-shaped window is assumed to be constant as the peak position changes, which is difficult to achieve. Future improvements to this work could be made by characterizing and mitigating changes in the bell-shaped window.


\begin{figure}
	\centering
	\includegraphics[width=6.6cm,height=4.8cm]{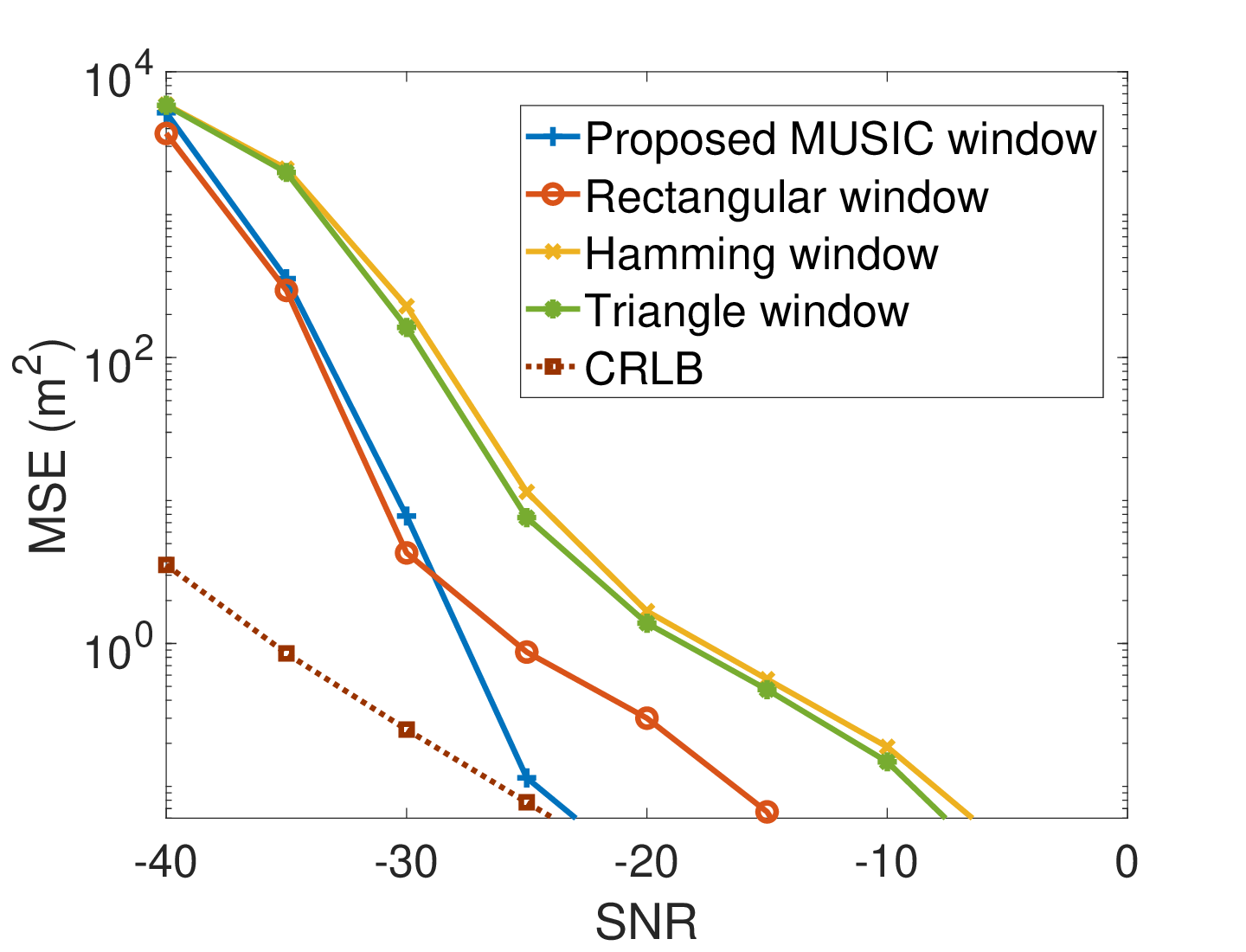}
	\caption{The synchronization MSE of different windows and the bell-shaped window. 
	}\label{fig10}
\end{figure}

In Fig. \ref{fig10}, we evaluate the synchronization performance of different windows by Monte Carlo simulations. As shown in Fig. \ref{fig10}, it is evident that the rectangular window has the best performance among the traditional windows, while the bell-shaped window outperforms the rectangular window. This phenomenon conforms to the criterion. However, the performance of the bell-shaped window is almost the same with that of the rectangular window, when the SNR is less than $-30$ dB. This phenomenon appears, since  the noise floor of the bell-shaped window increases with the SNR decreasing and consequently the peak location performance becomes worse.  

\section{Conclusions}
Window functions  designed have been investigated for fingerprint-spectrum-based passive sensing tailored to PMNs. Concretely, we derived the expressions of an  asymptotically optimal window and of the corresponding synchronization MSE. However, the asymptotically optimal window cannot be practically harnessed since it is strongly dependent on the specific time delays and gains corresponding to each propagation path, which are unknown before successful sensing. To circumvent this issue, we proposed a practical window selection criterion, and numerical simulations have verified our analysis and the criterion. 

\bibliographystyle{IEEEtran}
\bibliography{IEEEabrv,reference.bib}

\begin{thebibliography}{10}
\providecommand{\url}[1]{#1}
\csname url@samestyle\endcsname
\providecommand{\newblock}{\relax}
\providecommand{\bibinfo}[2]{#2}
\providecommand{\BIBentrySTDinterwordspacing}{\spaceskip=0pt\relax}
\providecommand{\BIBentryALTinterwordstretchfactor}{4}
\providecommand{\BIBentryALTinterwordspacing}{\spaceskip=\fontdimen2\font plus
\BIBentryALTinterwordstretchfactor\fontdimen3\font minus
  \fontdimen4\font\relax}
\providecommand{\BIBforeignlanguage}[2]{{%
\expandafter\ifx\csname l@#1\endcsname\relax
\typeout{** WARNING: IEEEtran.bst: No hyphenation pattern has been}%
\typeout{** loaded for the language `#1'. Using the pattern for}%
\typeout{** the default language instead.}%
\else
\language=\csname l@#1\endcsname
\fi
#2}}
\providecommand{\BIBdecl}{\relax}
\BIBdecl

\bibitem{9858656}
K.~Meng, Q.~Wu, S.~Ma, W.~Chen, K.~Wang, and J.~Li, ``Throughput maximization
  for {UAV}-enabled integrated periodic sensing and communication,'' \emph{IEEE
  Transactions on Wireless Communications}, vol.~22, no.~1, pp. 671--687, Jan.
  2023.

\bibitem{10226306}
K.~Meng, Q.~Wu, W.~Chen, and D.~Li, ``Sensing-assisted communication in
  vehicular networks with intelligent surface,'' \emph{IEEE Transactions on
  Vehicular Technology}, vol.~73, no.~1, pp. 876--893, Jan. 2024.

\bibitem{9737357}
F.~Liu, Y.~Cui, C.~Masouros, J.~Xu, T.~X. Han, Y.~C. Eldar, and S.~Buzzi,
  ``Integrated sensing and communications: Toward dual-functional wireless
  networks for {6G} and beyond,'' \emph{IEEE Journal on Selected Areas in
  Communications}, vol.~40, no.~6, pp. 1728--1767, Jun. 2022.

\bibitem{9868166}
Y.~Cao, S.~Yang, Z.~Feng, L.~Wang, and L.~Hanzo, ``Distributed spatio-temporal
  information based cooperative {3D} positioning in {GNSS}-denied
  environments,'' \emph{IEEE Transactions on Vehicular Technology}, vol.~72,
  no.~1, pp. 1285--1290, Jan. 2023.

\bibitem{zhang2021enabling}
J.~A. Zhang, M.~L. Rahman, K.~Wu, X.~Huang, Y.~J. Guo, S.~Chen, and J.~Yuan,
  ``Enabling joint communication and radar sensing in mobile networks-{A}
  survey,'' \emph{IEEE Communications Surveys \& Tutorials}, vol.~24, no.~1,
  pp. 306--345, First quarter. 2021.

\bibitem{10091198}
X.-Y. Wang, S.~Yang, T.-H. Yuan, H.-Y. Zhai, J.~Zhang, and L.~Hanzo,
  ``High-performance low-complexity hierarchical frequency synchronization for
  distributed massive {MIMO-OFDMA} systems,'' \emph{IEEE Transactions on
  Vehicular Technology}, vol.~72, no.~9, pp. 12\,343--12\,348, Sep. 2023.

\bibitem{ni2021uplink}
Z.~Ni, J.~A. Zhang, X.~Huang, K.~Yang, and J.~Yuan, ``Uplink sensing in
  perceptive mobile networks with asynchronous transceivers,'' \emph{IEEE
  Transactions on Signal Processing}, vol.~69, pp. 1287--1300, Feb. 2021.

\bibitem{IndoTrack}
X.~Li, D.~Zhang, Q.~Lv, J.~Xiong, S.~Li, Y.~Zhang, and H.~Mei, ``{IndoTrack}:
  Device-free indoor human tracking with commodity {Wi-Fi},'' \emph{Proceedings
  of the ACM on Interactive, Mobile, Wearable and Ubiquitous Technologies},
  vol.~1, no.~3, p. 1–22, Sep. 2017.

\bibitem{FarSense}
Y.~Zeng, D.~Wu, J.~Xiong, E.~Yi, R.~Gao, and D.~Zhang, ``{FarSense}: Pushing
  the range limit of {WiFi}-based respiration sensing with {CSI} ratio of two
  antennas,'' \emph{Proceedings of the ACM on Interactive, Mobile, Wearable and
  Ubiquitous Technologies}, vol.~3, no.~3, pp. 1--26, Sep. 2019.

\bibitem{zeng2020multisense}
Y.~Zeng, D.~Wu, J.~Xiong, J.~Liu, Z.~Liu, and D.~Zhang, ``{MultiSense}:
  Enabling multi-person respiration sensing with commodity {WiFi},''
  \emph{Proceedings of the ACM on Interactive, Mobile, Wearable and Ubiquitous
  Technologies}, vol.~4, no.~3, pp. 1--29, Sep. 2020.

\bibitem{9904500}
X.~Li, J.~A. Zhang, K.~Wu, Y.~Cui, and X.~Jing, ``{CSI}-ratio-based doppler
  frequency estimation in integrated sensing and communications,'' \emph{IEEE
  Sensors Journal}, vol.~22, no.~21, pp. 20\,886--20\,895, Nov 2022.

\bibitem{10616023}
W.~Jiang, Z.~Wei, S.~Yang, Z.~Feng, and P.~Zhang, ``Cooperation based joint
  active and passive sensing with asynchronous transceivers for perceptive
  mobile networks,'' \emph{IEEE Transactions on Wireless Communications}, Jul.
  2024, early access.

\bibitem{wxy}
X.-Y. Wang, S.~Yang, J.~Zhang, C.~Masouros, and P.~Zhang, ``Clutter
  suppression, time-frequency synchronization, and sensing parameter
  association in asynchronous perceptive vehicular networks,'' \emph{IEEE
  Journal on Selected Areas in Communications}, Aug. 2024, early access.

\bibitem{9921271}
Z.~Wei, Y.~Wang, L.~Ma, S.~Yang, Z.~Feng, C.~Pan, Q.~Zhang, Y.~Wang, H.~Wu, and
  P.~Zhang, ``{5G PRS}-based sensing: A sensing reference signal approach for
  joint sensing and communication system,'' \emph{IEEE Transactions on
  Vehicular Technology}, vol.~72, no.~3, pp. 3250--3263, Mar. 2023.

\bibitem{liu2020super}
Y.~Liu, G.~Liao, Y.~Chen, J.~Xu, and Y.~Yin, ``Super-resolution range and
  velocity estimations with {OFDM} integrated radar and communications
  waveform,'' \emph{IEEE Transactions on Vehicular Technology}, vol.~69,
  no.~10, pp. 11\,659--11\,672, Oct. 2020.

\bibitem{10640151}
X.-Y. Wang, S.~Yang, H.-Y. Zhai, C.~Masouros, and J.~Andrew~Zhang, ``Windowing
  optimization for fingerprint-spectrum-based passive sensing in perceptive
  mobile networks,'' \emph{IEEE Transactions on Communications}, Aug. 2024,
  early access.

\bibitem{6942180}
A.~Naha, A.~K. Samanta, A.~Routray, and A.~K. Deb, ``Determining
  autocorrelation matrix size and sampling frequency for {MUSIC} algorithm,''
  \emph{IEEE Signal Processing Letters}, vol.~22, no.~8, pp. 1016--1020, Aug.
  2015.

\end{thebibliography}
\end{document}